\documentclass[a4paper,fleqn,usenatbib]{mnras}

\UseRawInputEncoding

\newcommand{\msun}{$M_\odot$}
\newcommand{\rsun}{$R_\odot$}
\newcommand{\lsun}{$L_\odot$}
\newcommand{\kms}{\,km\,s$^{-1}$}

\usepackage[T1]{fontenc}
\usepackage{ae,aecompl}

\usepackage{graphicx}	
\usepackage{amsmath}	
\usepackage{amssymb}	

\title[The symbiotic recurrent nova V3890~Sgr]{The symbiotic recurrent nova V3890~Sgr: binary parameters and pre-outburst activity}

\author[J. Miko{\l}ajewska et al.]{
J. Miko{\l}ajewska$^{1}$\thanks{E-mail: mikolaj@camk.edu.pl},
K. I{\l}kiewicz$^{2}$,
C. Ga{\l}an$^{1}$,
B. Monard$^{3}$,
M. Otulakowska-Hypka$^{4}$,
M.~M. Shara$^{5}$,
\newauthor
and A. Udalski$^{6}$
\\
$^{1}$Nicolaus Copernicus Astronomical Center, Polish Academy of Sciences, Bartycka 18, 00716 Warsaw, Poland \\
$^{2}$Centre for Extragalactic Astronomy, Department of Physics, University of Durham, South Road, Durham DH1 3LE, UK \\
$^{3}$Kleinkaroo Observatory, Calitzdorp, Western Cape, South Africa\\
$^{4}$Astronomical Observatory Institute, Faculty of Physics, Adam Mickiewicz University, S{\l}oneczna 36, 60286 Pozna{\'n}, Poland\\
$^{5}$Department of Astrophysics, American Museum of Natural History, Central Park West at 79th Street, New York, NY 10024, USA\\
$^{6}$Astronomical Observatory, University of Warsaw, Al. Ujazdowskie 4, 00478 Warsaw, Poland\\
}

\date{Accepted XXX. Received YYY; in original form ZZZ}

\pubyear{2020}

\begin{document}
\label{firstpage}
\pagerange{\pageref{firstpage}--\pageref{lastpage}}
\maketitle

\begin{abstract}
We present and analyze optical photometry and high resolution SALT spectra of the symbiotic recurrent nova V3890 Sgr at quiescence.  The orbital period, $P=747.6$ days has been derived from both photometric and spectroscopic data. Our double-line spectroscopic orbits indicate that the mass ratio is $q=M_{\rm g}/M_{\rm WD} = 0.78 \pm 0.05$, and that the component masses are $M_{\rm WD}  \approx 1.35\pm0.13$\,\msun\, and  $M_{\rm g} \approx 1.05 \pm 0.11$\,\msun. The orbit inclination is $\approx 67-69\degr$. The red giant is filling (or nearly filling) its Roche lobe, and the distance set by its Roche lobe radius, $d \approx 9$ kpc, is consistent with that resulting from the giant pulsation period. The outburst magnitude of V3890 Sgr is then very similar to those of RNe in the Large Magellanic Cloud. 
V3890 Sgr shows remarkable photometric and spectroscopic activity between the nova eruptions with timescales similar to those observed in the symbiotic recurrent novae T CrB and RS Oph and Z And-type symbiotic systems. 
The  active source has a double-temperature structure which we have associated with the presence of an accretion disc. The activity would be then caused by changes in the accretion rate. 
We also provide evidence that V3890 Sgr contains a CO WD accreting at a high, $\sim$ a\,few$\times 10^{-8}$--$ 10^{-7}$\,\msun\, yr$^{-1}$,  rate. The WD is growing in mass, and should give rise to a Type Ia supernova within $\la 10^6$ yrs - the expected lifetime of the red giant.
\end{abstract}




\section{Introduction}

Cataclysmic variables (CVs) are among the most common interacting binaries in which a red/brown dwarf donor (the majority of CVs), or a subgiant or giant donor transfers material to a white dwarf (WD) companion. The accretion leads to various outburst phenomena, of which thermonuclear nova explosions are the most spectacular. As the H-rich matter deposited on the WD surface increases in mass, the pressure at its base increases and eventually the temperature becomes high enough to trigger a thermonuclear runaway (TNR) reaction. When this happens the envelope is explosively blown off the WD surface, and the binary becomes extremely luminous -- a nova. 

Nova explosions provide large enrichments to the interstellar medium in CNO as well as Ne, Na, Al and other intermediate-mass elements, and their study is essential to understanding Galactic nucleosynthesis including the 'life' elements. Furthermore, CO novae (i.e. those occurring on CO WDs) are the main 7Li factories in galaxies and they can produce all the Li observed in excess of that predicted by Big Bang nucleosynthesis \citep[e.g.,][]{molaro2016}.

Whereas all TNR novae are recurrent with timescales from months \citep{darnley2014} to millions of years \citep[e.g.,][]{yaron2005}, there are only a few dozen novae with two or more TNR explosions recorded in the Milky Way and nearby galaxies; only these are called recurrent novae (RNe). RNe must contain WDs close to the Chandrasekhar limit, and they must be accreting at very high rates $\sim 1-7 \times 10^{-7}\, M\sun\, \rm yr^{-1}$ \citep{kato1991,yaron2005,hil16} in order to build critical-mass envelopes quickly enough to erupt so frequently. Four of ten known Galactic RNe have red giant donors: T CrB, V745 Sco, RS Oph, and V3890 Sgr. These are also known as symbiotic RNe (SyRNe) because of their close relation with symbiotic stars, interacting binaries in which an evolved giant transfers material to a hot and luminous companion \citep[for recent reviews see, e.g.,][]{anupama2013,mik2013}.

Understanding classical novae (CNe), and in particular, RNe is also essential to solve one of the most pressing problems in modern astrophysics -- what are the progenitors of Type Ia supernovae (SNIa)? There is general consensus that they result from thermonuclear disruption of a CO WD reaching the Chandrasekhar mass either due to mass accretion from a non-degenerate donor (single-degenerate or SD model) or merger of two WDs (double-degenerate or DD scenario) \citep[see,][and references therein]{maoz}. 

RNe with their massive WDs and high accretion rates are promising candidates for the SD channel provided that these WDs can grow secularly in mass to approach the Chandrasekhar limit and ignite carbon burning to produce an SNIa. Since CO WDs are not born with masses in excess of $1.1\, M\sun$ \citep{ritossa1996}, and CNe eject most of their accreted and burnt mass, it may be very difficult, if not impossible for CNe to produce SNIa. 

The situation looks better for SyRNe. Our recent determination that the WD in RS Oph is composed of CO and its mass has grown significantly since its birth makes RS Oph (and, in general, SyRNe) a prime candidate for a future SNIa via the SD scenario \citep{MS2017}. Recent theoretical simulations show that the SD symbiotic channel could produce SNe Ia with intermediate and old ages, contributing up to 5\,\% of all SNIa in the Galaxy \citep{liu2019}. Moreover, SNIa can result even from the widest symbiotic systems with a massive WD and Mira donor like V407 Cyg \citep{ilkiewicz2019}.

In this study, we present and discuss the spectroscopic and photometric behaviours of V3890 Sgr at quiescence. We review the current state of knowledge of V3890 in Section~\ref{V3890}, then describe our data in Section~\ref{obs}.  The binary parameters of this symbiotic nova, derived from our double-line spectroscopic orbits are presented in Section~\ref{parameters}. We characterize and discuss the activity of V3890 Sgr prior to its TNR nova outburst in Section~\ref{activity}. Our results are discussed in the context of the SNIa in Section~\ref{snia}, and briefly summarized in Section~\ref{conclusions}.

\section{V3890 Sgr}\label{V3890}

The first known eruption of V3890 Sgr occurred in 1962; its detection was published a decade later by \citet{din73}. The star brightened by at least seven magnitudes between 10 May and 2 June 1962. It faded $\sim$3 magnitudes in $\sim$ 13 days, and returned to quiescence on a two months timescale. A second eruption 28 years later, on 27 April 1990 \citep{kil90} established V3890 Sgr as one of the ten known Galactic RNe. 
Early optical and infrared spectra \citep{wag90,har93} of the 1990 outburst showed strong and broad emission lines of \ion{H}{i} and \ion{He}{i}.
The bright infrared \ion{H}{i} lines had an  average FWHM of  1500\,\kms \citep{har93} significantly lower than the expansion velocity of 2140\,\kms estimated from the FWHM of H$\alpha$ \citep{wag90}.
Low-dispersion IUE observations of V3890 Sgr 18 days after its 1990 eruption showed strong emission lines of \ion{N}{v}, \ion{O}{i}, \ion{Si}{iv}, \ion{N}{iv}], \ion{C}{iv}, \ion{He}{ii}, \ion{O}{iii}], \ion{N}{iii}], \ion{Si}{iii}], \ion{C}{iii}], and \ion{Mg}{ii} \citep{gon90}.
Most of the lines had a broad (FWHM $\sim 4000$ \kms) component and a narrower core (FWHM $\sim 1200$ \kms, comparable to the spectral resolution used). However, we should note that all these lines are blends, and the expansion velocities may be in fact lower than the values indicated by their FWHMs.
Within two weeks the ionization level rose as emission lines of [\ion{Fe}{xiv}] and [\ion{Ar}{x}] appeared \citep{muk90}. A month later the red continuum and TiO absorption lines of an M giant appeared as the nova dimmed, and the FWHM of the emission lines dropped to a few hundred \kms.
\citet{wil91} classified the secondary spectral type as M8 III soon after the 1990 outburst, while weaker TiO bands seen in 1997 and 1998 led \citet{anupama1999} to classify the secondary in the range K5.5-M6 III. 

\citet{anupama1999} demonstrated that V3890 Sgr is photometrically and spectroscopically variable at quiescence like all SyRNe in their study.  They also associated this behaviour with the presence of a disc-accreting WD embedded in the optically thick wind from its giant companion.

\citet{schaefer2009} was the first to report an orbital period for V3890 Sgr: 519.7$\pm$0.3 days, based on claimed ellipsoidal light variations and shallow eclipse. However, \citet{mroz2014} could not confirm the orbital period nor the ellipsoidal variability in the OGLE light curves.
The light curves of V3890 Sgr since 1900 were also collected and discussed by \citet{schaefer2010} in his comprehensive photometric histories of all Galactic RNe. 

V3890 Sgr's third nova outburst began on 27 August 2019.  The X-ray light curve and spectra \citep{pag20,orio2020,sin21} showed initial hard shock emission, supersoft emission with a peak blackbody temperature of 100 eV beginning $\sim$ 8.5 days later, and ending by day 26 of the outburst. These timings suggest a WD mass $\sim 1.3$\,\msun.

\section{Observations}\label{obs}

\subsection{Photometry}

Photometric monitoring of  V3890~Sgr has been carried out with  a 35cm Meade RCX400 telescope in Kleinkaroo Observatory using  a SBIG ST8-XME CCD camera and  $V$ and $Ic$ filters. Each observation was the result of several individual exposures, which were calibrated (dark-subtraction and flat-fielding) and stacked selectively. Magnitudes were derived from differential photometry to nearby reference stars using the single image mode of AIP4 image processing software. The photometric accuracy of the derived magnitudes is better than 0.1 mag. Additional $V, Ic$ photometric data in this study are from the OGLE project (\citealt{mroz2014}) updated to include recent OGLE-VI measurements. The typical error of measured magnitudes in this survey was 0.002 mag. 

We have also retrieved publicly available $m_{\rm pg}/B, V, R, Ic, J$ photometry collected by \citet{schaefer2010}.

\subsection{Spectroscopy}

Spectroscopic observations were carried out in 2014-2020 with the  11m 
Southern African Large Telescope (SALT; Buckley, Swart \& Meiring 2006; 
O'Donoghue et al. 2006) using the High Resolution Spectrograph 
(HRS; Bramall et al. 2010, 2012; Crause et al. 2014). 
The exposure times were 2400~s.
The spectrograph was used in a medium resolution (MR) mode with resolving power R$\sim$40000 
and wavelength coverage of 4000 -- 8800\,\AA. 
The initial data reduction was performed using the pysalt pipeline (Crawford et al. 2010) 
which was followed by the HRS pipeline (Kniazev, Gvaramadze \& Berdnikov 2016), 
based on the MIDAS feros (Stahl, Kaufer \& Tubbesing 1999) and echelle (Ballester 1992) packages. 

Low-resolution  spectra were also observed around the \ion{Ca}{ii} triplet ($\lambda \sim 7200-9600$\,\AA, R $\sim$ 2200) as well as several M-type standards with the SpUpNIC spectrograph (grating nr 11) mounted on the 1.9m `Radcliffe' telescope at the South African Astronomical Observatory. The observations were obtained in 2017, October 28 as part of our project (JM, CG) to look for chemical peculiarities of symbiotic giants. These spectra are used here to improve the spectral classification of the M giant.

\begin{figure}
\includegraphics[width=\columnwidth]{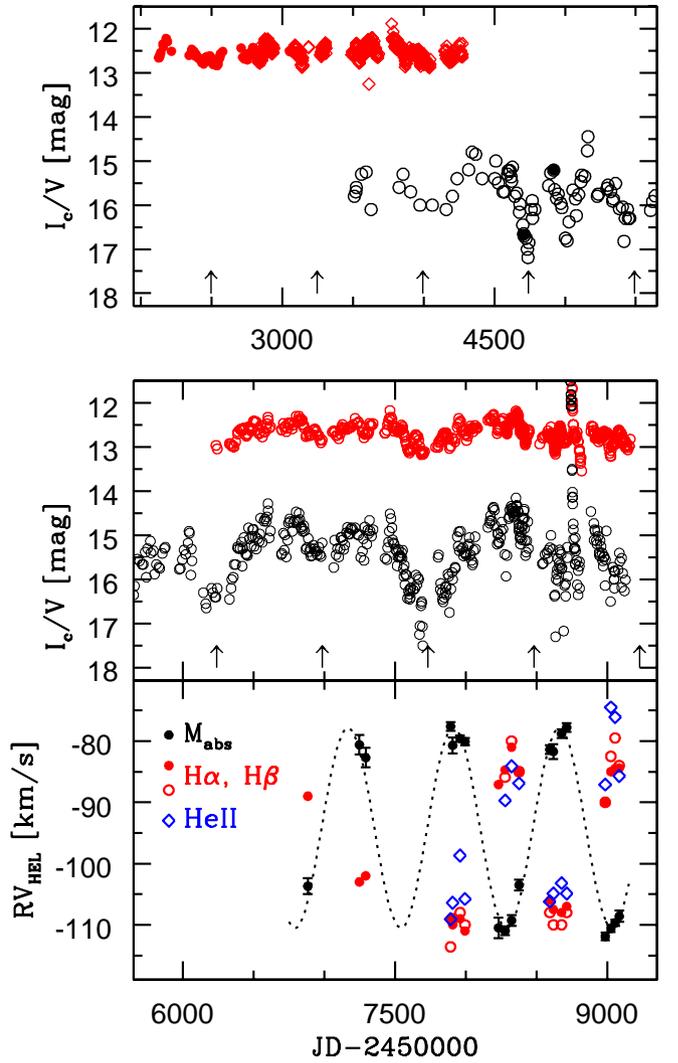}
\caption{({\it Top/middle}) $V$ (black)  and $I_{c}$ (red) light curves of V3890 Sgr in 2001-2020. Open and filled circles correspond to the Kleinkaroo and OGLE photometry, respectively, whereas diamonds  represent the data from \citet{schaefer2010}. The zero-points of the Kleinkaroo and Schaefer's $I_{c}$ photometry are shifted by $\Delta I 
=0.24$ and $-0.40$ mag, respectively, to fit the OGLE light curve.
Arrows mark times of minima given by Eq.~\ref{ephemeris}.
({\it Bottom}) Radial velocity data for the red giant (black dots), and the emission line wings of H$\alpha$ wings (red dots), H$\beta$ (red open circles) and \ion{He}{ii}\,4686 (blue diamonds). The dotted line represents the circular orbit solution from Table~\ref{orbits}.  }\label{var}
\end{figure}

\section{Binary parameters}\label{parameters}

The light curves of V3890 Sgr (Fig.~\ref{var}) reveal, in addition to the 104.5-day pulsations aforementioned by \citet{mroz2014}, remarkable and complex variability on timescales of months to years. Some of these variations seem to be the result of intrinsic variability of the hot companion as also indicated by our SALT spectra of V3890 Sgr. In general, they show a late M-star continuum, most apparent longwards of $\lambda \ga 7200$\,\AA\, superposed with strong emission lines, mainly of \ion{H}{i}, \ion{He}{i}, \ion{Fe}{ii} as well as highly ionized \ion{He}{ii} and  [\ion{O}{iii}] lines indicating significant activity of V3890 Sgr during our observations. Detailed discussion of this activity is in Sec.~\ref{activity}.

\subsection{Orbital variability and spectroscopic orbits}\label{sp_orbits}

We have analysed the $V$ and $I_{c}$ light curves (shown in Fig.~\ref{var}) using Lomb-Scargle (LS) periodograms as well as the phase dispersion minimization \citep[PDM]{pdm} method which is better suited for light curves with big gaps. First of all, we could not see the 519.7-day orbital period reported by \citet{schaefer2009, schaefer2010}. Instead, we have found a $\sim 750$-day periodicity in both $V$ and $I_{c}$ light curves which we attribute to the orbital period. Our analysis of the $I_{c}$ data also confirms the pulsations with $P_{\rm pul}=104.5$ days detected by \citet{mroz2014}. 
We have also examined the  $m_{\rm pg}/B$ photometry collected by \citet{schaefer2010}. In all cases, the most regular changes have been obtained with $\sim 750$-day period, and the linear photometric ephemeris:

\begin{equation}
 JD_{min} = (2456985.2 \pm 5.8) + (747.60 \pm 0.33) \times E
 \label{ephemeris}
\end{equation}

\begin{figure}
\includegraphics[width=\columnwidth]{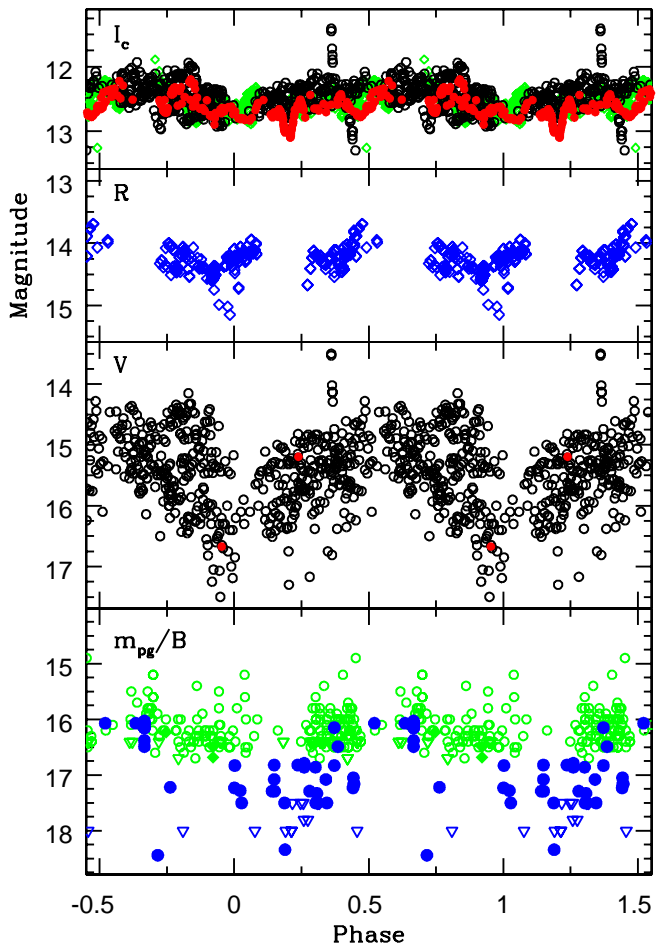}
\caption{Light cuves phased with the orbital ephemeris (Eq.~\ref{ephemeris}) in various filters. Black and red symbols correspond to the Kleinkaroo and OGLE photometry, respectively, whereas green and blue symbols  represent the Harvard College Observatory and Maria Mitchell Observatory photographic photometry (with upper limits shown as triangles), respectively, $R$ and $I$ photometry from \citet{schaefer2010}. }\label{lc}
\end{figure}

Fig.~\ref{lc} presents phase plots of the data. In general, the minimum is deeper at shorter wavelengths presumably due to a larger contribution of the hot companion. There is no evidence for eclipses although the most suitable light curve to look for eclipses $m_{\rm pg}/B$ is dominated by noise, and often only lower limits are available. However, the depth of minimum in $V$ light varies by up to 1 mag from cycle to cycle indicating that there are no total eclipses in the system. In the red/near infrared the red giant dominates, and its pulsation amplitude becomes comparable ($R$) or sometimes even larger ($I_{c}$) than orbital variations.

To derive the radial velocity curve of the red giant we have measured M-type absorption features  in the region $\lambda \sim 8000-8200$\,\AA\, corresponding to \ion{Fe}{i}, \ion{Na}{i}, \ion{V}{i}.  

We have also identified and measured absorption of singly ionized elements in the blue region of our SALT spectra ($\lambda \sim 4400-4600$\,\AA), corresponding to \ion{Cr}{ii}, \ion{Fe}{ii}, \ion{Ti}{ii}, which resemble those of an F-G supergiant. A similar set of absorption lines was identified in quiescent spectra of another symbiotic recurrent nova RS Oph \citep[][and references therein]{brandi2009}. This so called cF-shell absorption system has also been found in other active symbiotic systems, and it is believed to be linked to the hot component and/or the material streaming towards the hot component \citep[e.g.][2009, and references therein]{mk1992, quiroga2002, brandi2005}.

\begin{figure}
\includegraphics[width=\columnwidth]{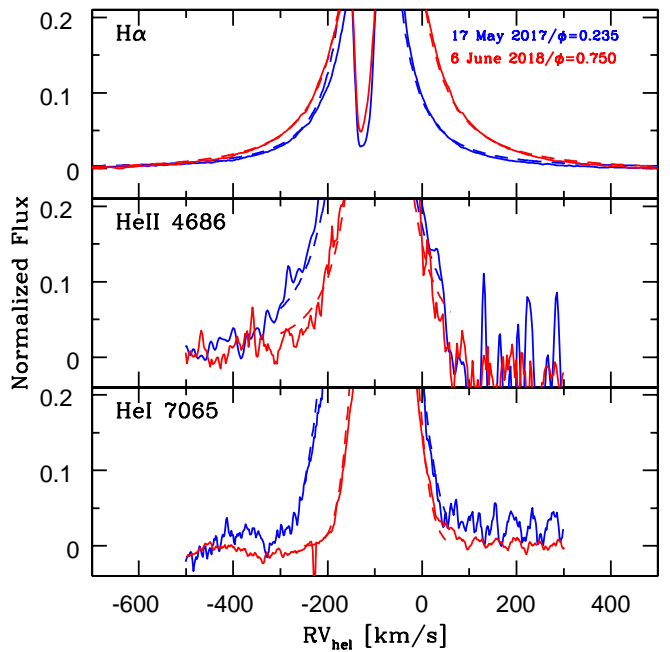}
\caption{Examples of the Gaussian (\ion{He}{i} 7065) and Lorentz (\ion{He}{ii}\,4686 and H$\alpha$) profile fit (dashed lines) of the emission line wings at different orbital phases. The noise on the red side of \ion{He}{ii} line is due to imperfect merging of the adjacent spectral orders. }\label{fits}
\end{figure}

\begin{figure}
\includegraphics[width=\columnwidth]{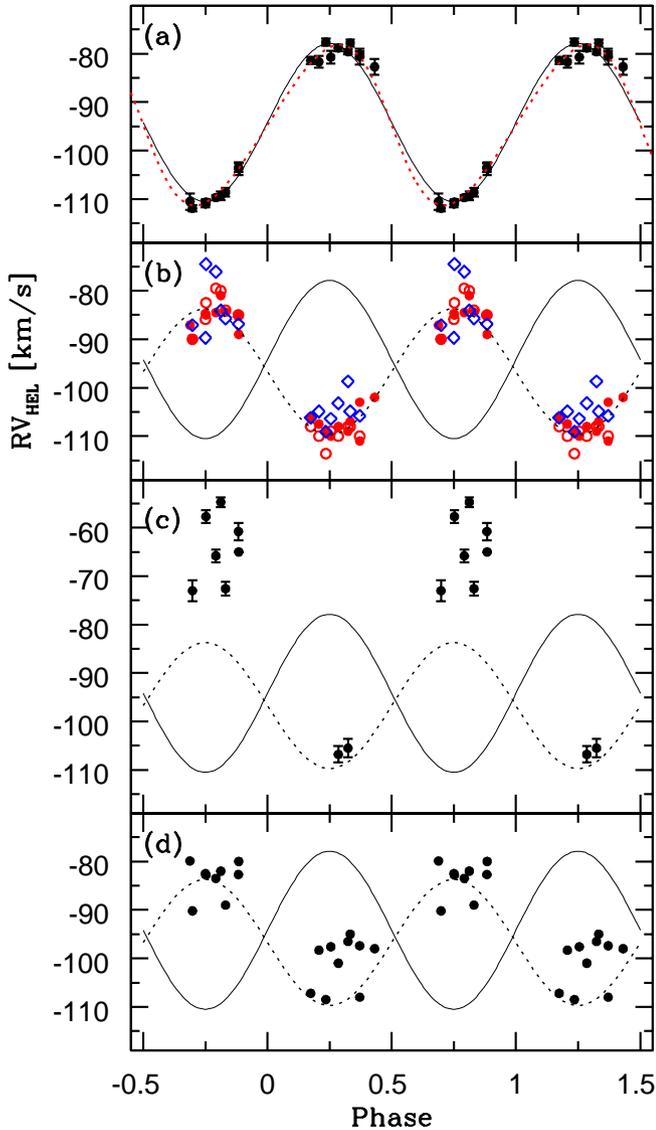}
\caption{Radial velocity data and orbital solutions for V3890 Sgr: (a)  M-giant absorption lines, (b) H$\alpha$ (red filled circles), H$\beta$ (red open circles) and \ion{He}{ii}\,4686 (blue diamonds) wings, (c) cF-absorption system and (d) \ion{He}{i} emission triplet lines. The black solid and dotted lines repeat the circular orbit of the M-giant and the \ion{H}{i} emission line wings, respectively. The red dotted line represents  the best elliptical fit to the M-giant absorption lines ($e=0.13$, $\omega=90\degr$).}\label{rvel}
\end{figure}

Individual radial velocities were obtained by a Gaussian fit of the line profiles, and  a mean value was calculated for each spectrum. The resulting heliocentric velocities together with their standard deviations ($\sigma$) are given in Table~\ref{tab_rvel} and plotted in Fig.~\ref{var}.

Since the broad emission wings of \ion{H}{i} and \ion{He}{ii} reflect the orbital motion in a number of symbiotic stars, and the symbiotic recurrent novae RS Oph \citep[e.g.][2009, and references therein]{quiroga2002,brandi2005} and T CrB \citep{stanishev2004}, we have also derived their radial velocities by fitting both Gaussian and Lorentz profiles to the wings. We fitted only the outer parts of the line that are not contaminated by the central absorption(s) and asymmetry. Furthermore, we have applied the same method to estimate the radial velocities of the strongest \ion{He}{i} line wings. 
Examples of these fits are shown in Fig.~\ref{fits}. Typical errors in individual radial velocity measurements are 2--3 \kms\ for the \ion{H}{i} emission wings, and 3--5 \kms\ in the case of \ion{He}{ii} and \ion{He}{i}. The red wing of the \ion{He}{ii}\,4686 line in some spectra is affected by its location in the overlapping region of adjacent spectral orders (see Fig.~\ref{fits}).
The heliocentric radial velocities for the emission line wings are also given in Table~\ref{tab_rvel} and shown in Fig.~\ref{var}.

The radial velocities of the cool giant and the broad emission wings (Fig.~\ref{var}) all show the same $\sim 750$-day periodicity. The emission line wings clearly vary in anti-phase with the M-giant absorptions. Moreover, the inflection points of the M-giant radial velocity curve coincide with the photometric minima whereas those of the emission features coincide with the photometric maxima as expected for the orbital motion.

Table~\ref{orbits} lists the orbital solutions for the M-giant with  a weight $w_i = \sigma_i^{-1}$ applied to the data. However, we noted that the weighted solutions show no important differences when a weight 1 is applied to all data.
An elliptical orbit ($e=0.13\pm0.03$) fits the radial velocity curve slightly better than the circular one (Fig~\ref{rvel}). In both cases, the orbital period, $P_{\rm orb} = 740\pm8$ and $P_{\rm orb} = 747\pm4$ days for the circular and elliptical solution, respectively, as well as the time of inferior conjunction agree very well with the photometric ephemeris (Eq.~\ref{ephemeris}). We have therefore adopted $P_{\rm orb} = 747.6$ days for the final solutions. Moreover, the periastron longitude of the elliptical orbit is near to $90\degr$  indicating that the apparent eccentricity may be an effect tied to the line of sight, and be due to tidal distortion of the giant. For the final elliptical orbit solution  $\omega=90\degr$ was assumed.
For the emission line wings only circular solutions with fixed period are included in Table~\ref{orbits} because they provide the best fit to the data.
These solutions show clear anti-phase changes of both radial velocity sets and the systemic velocity agrees very well with that of the M-giant (Fig~\ref{rvel}). 

The solutions for the \ion{H}{i} and \ion{He}{ii} 
emission wings agree within their respective errors which suggests that both are formed in the same region near the WD. Since the measurements for the 
\ion{He}{ii} radial velocities and the resulting orbital solution are much less accurate than those for \ion{H}{i} lines, we have decided to use only the  \ion{H}{i} lines for the component mass determination.
The \ion{He}{i} lines give distinctly lower semi-amplitude than both \ion{H}{i} and \ion{He}{ii} emission wings. Such reduced amplitude can be explained by a contribution from material between the two components. Similar reduced amplitude of \ion{He}{i} lines with respect to the hot component is observed in other symbiotic stars, and these lines do not seem to be a good tracer of the WD \citep[e.g.,][] {quiroga2002}.

\begin{table*}
 \centering
  \caption{Heliocentric radial velocities of the red giant absorption lines, the blue cF-type absorption lines,  and emission wings of \ion{H}{i}, \ion{He}{ii} and \ion{He}{i} lines in V3890 Sgr.}\label{tab_rvel}
  \begin{tabular}{@{}lccccccccc@{}}
  \hline
Date & JD &  Phase & M-abs &  cF-abs &  \ion{H}{i} 6562.817 & \ion{H}{i} 4861.332 & \ion{He}{ii}\,4685.68 & \ion{He}{i} (t)$^1$ & \ion{He}{i} (s)$^2$\cr
 &  2400000+ &  &  \multicolumn{7}{c}{$RV\,[\rm km\,s^{-1}]$}\cr
\hline
12 Aug 2014 & 56882.4 & 0.884 & -103.7$\pm$1.3 & -65.0$\pm0.2^3$ & -89 &  & & -80.0$\pm1.2$ & -86\\
11 Aug 2015 & 57246.4 & 0.371 & -80.6$\pm$1.6 &  & -103 & & & -108$\pm3.0$ & -103\\
25 Sep 2015 & 57291.3 & 0.431 & -82.7$\pm$1.6 &  & -102 & &  & -98.0$\pm1.0$ & -101 \\
17 May 2017 & 57891.4 & 0.235 & -77.6$\pm$0.7 &  & -109 & -113.6 & -109.1 & -108.5$\pm1.5$ & -113\\
1 Jun 2017 &  57906.4 & 0.255 & -80.7$\pm$1.3 &  & -110 & -109.0 & -106.4 & -97.6$\pm1.2$ & -99\\
23 Jul 2017 & 57958.3 & 0.324 & -79.6$\pm$0.5 & -105.5 $\pm$1.9 & -109 & -108 & -98.7 & -96.5$\pm4.5$ & -107\\
27 Aug 2017 & 57993.4 & 0.371 & -80.1$\pm$0.6 &  & -111 & -110 & -105.8 & -97.4$\pm0.9$ & -111\\
20 Apr 2018 & 58229.5 & 0.688 & -110.5$\pm$ 1.7 &  & -87.1 & & & -79.9$\pm1.7$ & -92\\
6 Jun 2018 &  58276.4 & 0.750 & -111.0$\pm$0.7 &  & -84.7 & -85.9 & -89.7 & -82.5$\pm1.5$ & -87\\
22 Jul 2018 & 58322.3 & 0.812 & -109.3$\pm$0.9 & -54.7 $\pm$1.0 & -81.0 & -80.0 & -84.1 & -82.0$\pm5.8$ & -87\\
13 Sep 2018 & 58375.3 & 0.883 & -103.5$\pm$0.9 & -60.8 $\pm$1.8 & -85.1 & -85 & -86.9 & -82.7$\pm1.8$ & -94\\
18 Apr 2019 & 58592.5 & 0.174 & -81.4$\pm$0.7 &  & -106 & -108 & -106.2 & -107.2$\pm3.2$ & -114\\
13 May 2019 & 58617.4 & 0.207 & -81.7$\pm$1.2 &  & -107.5 & -110 & -104.9 & -98.3$\pm1.7$ & -103\\
10 Jul 2019 & 58675.3 & 0.285 & -78.8$\pm$0.7 & -106.8 $\pm$1.7 & -108 & -110 & -103.2 & -101$\pm2.5$ & -108\\
15 Aug 2019 & 58711.4 & 0.333 & -77.8$\pm$0.7 &  & -107 & -108 & -104.9 & -95.0$\pm2.2$ & -94\\
14 May 2020 & 58984.4 & 0.698 & -111.9$\pm$0.6 & -73.0 $\pm$2.2 & -90.0 & -90 & -87.1 & -90.2$\pm0.5$ & -95\\
23 Jun 2020 & 59024.3 & 0.752 & -110.6$\pm$0.6 & -57.7 $\pm$1.3 & -85.0 & -82.5 & -74.5 & -82.8$\pm2.2$ & -89\\
23 Jul 2020 & 59054.5 & 0.792 & -109.7$\pm$0.5 & -65.8 $\pm$1.3 & -84.5 & -79.5 & -76.1 & -83.5$\pm0.6$ & -99\\
21 Aug 2020 & 59083.4 & 0.831 & -108.6 $\pm$0.9 & -72.6 $\pm$1.4 & -84.4 & -84 & -85.7 & -89.0$\pm2.1$ & -102:\\
\hline
\end{tabular}
\begin{list}{}{}
\item $^{1}$ \ion{He}{i} 5875.64 and 7065.19 triplet lines.
\item $^{2}$ \ion{He}{i} 6678.15 singlet line.
\item $^3$  \ion{Na}{i} D lines.
\end{list}
\end{table*}

\begin{table*}
 \centering
  \caption{Orbital elements and related parameters of V3890 Sgr.}\label{orbits}
  \begin{tabular}{@{}lccccccccc@{}}
  \hline
Component & P [days]	 & $K\,[\rm km\,s^{-1}]$ & $V_0\,[\rm km\,s^{-1}]$ &	$e$ &  $\omega$ [deg] & $T_0^1$ & $\Delta T$ [days]$^2$ & $a\,\sin\,i $[AU] & $f(m)$[\msun]\\
\hline								
M-giant &	740$\pm$8 & 16.26 $\pm$0.46	& -94.25$\pm$0.40 & 0$^3$ & & 56990$\pm$15 & 5 & 1.103 &  0.330 \\
              & 747.6$^3$	&   16.31$\pm$0.45	& -94.19$\pm$0.39 &	0$^3$ &  & 56979$\pm$7 & -6& 1.117$\pm$0.031 & 0.336$\pm$0.028 \\
              & 747$\pm$4	& 16.52$\pm$ 0.29 &	-93.73$\pm$ 0.61 & 0.13$\pm$0.03 & 127$\pm$19 & 58922$\pm$ 41 & -7 & 1.122$\pm$0.030 &	0.341$\pm$0.024 \\
              & 747$^3$ &	16.62$\pm$0.38 & -94.72$\pm$0.71 & 0.13$\pm$0.03	& 90$^3$ & 58842$\pm$49 & -10 & 1.129$\pm$0.030	& 0.347$\pm$0.028 \\
\hline
Emission wings \cr							
H$\alpha$ \& H$\beta$ & 747.6$^3$ & 13.00$\pm$0.44 & -96.72$\pm$0.40 & 0$^3$ & & 56996$\pm$9 &	11 \\		
\ion{He}{ii}\,4685.68	 & 747.6$^3$ & 11.85$\pm$1.38	 & -94.06$\pm$1.17 & 0$^3$ & & 56969$\pm$32 & -16 \\		
\ion{He}{i} (t) & 747.6$^3$ & 9.72$\pm$1.43 & -92.60$\pm$1.22 & 0$^3$ & & 56977$\pm$34 & -8	\\	
\hline
\end{tabular}
\begin{list}{}{}
\item $^{1}$ $T_0$ - inferior conjuction/periastron passage for the circular/elliptic orbit, respectively.
\item $^{2}$ $\Delta T$  - time difference between  inferior conjuction and  photometric minimum (Eq.~\ref{ephemeris}).
\item $^3$  assumed.
\end{list}
\end{table*}

The semi-amplitudes of the M-giant and \ion{H}{i} emission wings indicate the mass ratio $q = M_{\rm g}/M_{\rm WD} = K_{\rm \ion{H}{i}}/K_{\rm g} = 0.78\pm0.05$, the component masses $M_{\rm g} \sin^3i = 0.86\pm0.05$ \,\msun\, and $M_{\rm WD} \sin^3i = 1.10\pm0.10$ \,\msun, and the binary separation $a \sin i=2.010 \pm 0.065$\,AU. Assuming that the white dwarf cannot exceed the Chandrasekhar limit ($M_{\rm WD} \loa 1.4$\,\msun), we estimate the lower limit for the orbit inclination, $i \goa 67\pm 4\degr$) whereas the upper limit, $i \la 69\degr$ is set by the absence of eclipses in the optical light curves.
This indicates the white dwarf in V3890 Sgr is very massive,  $M_{\rm WD} \goa 1.35\pm0.13$\,\msun.  The red giant mass is $M_{\rm g} \goa 1.05 \pm 0.11$\,\msun.

\begin{figure}
	\includegraphics[width=\columnwidth]{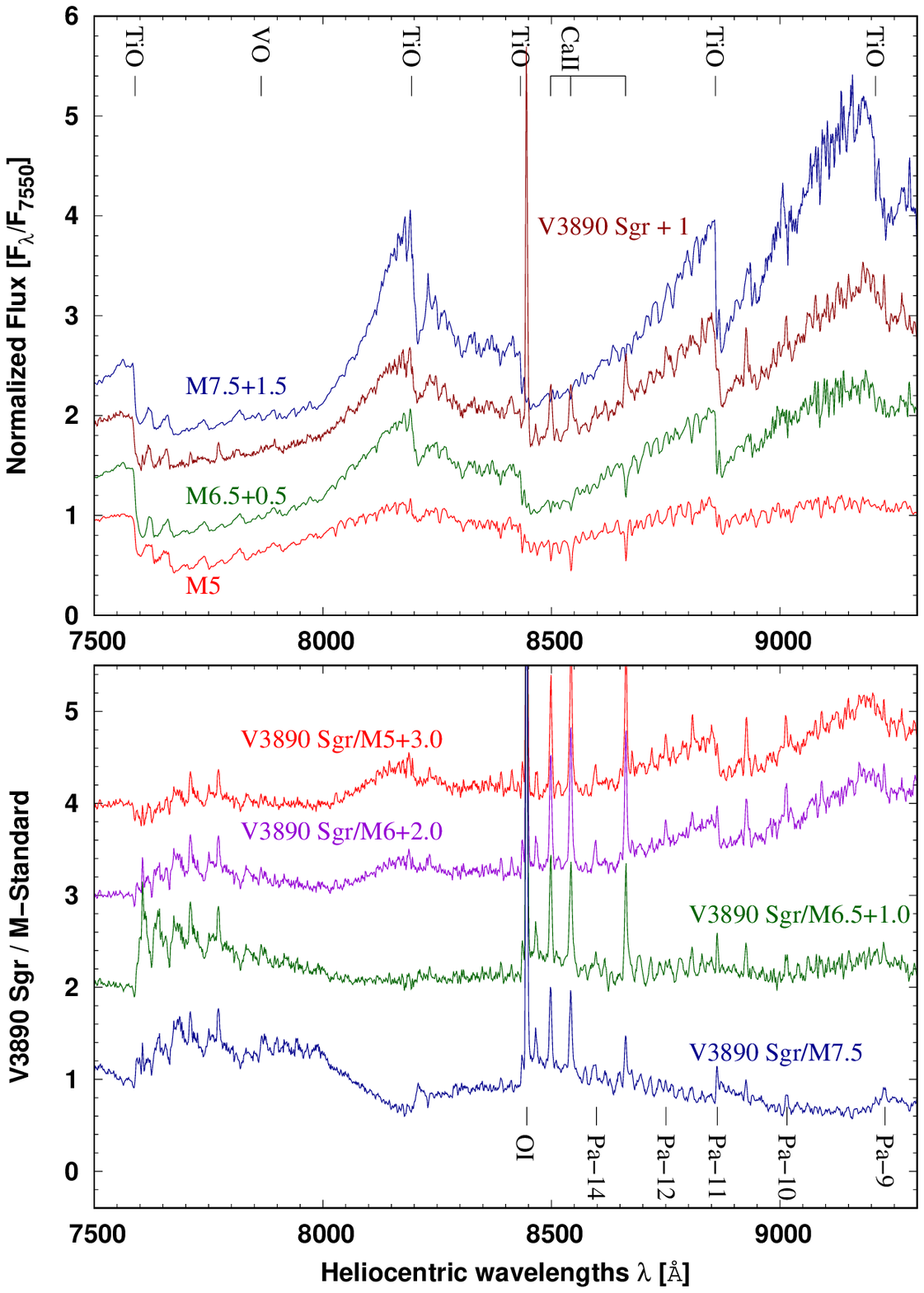}
    \caption{The spectrum of V3890 Sgr (dark red) compared with those of spectroscopic standards of spectral types ({\it top}) M7.5\,III, M6.5\,III and M5\,III. At the {\it bottom} the ratios of the V3890 Sgr spectrum to those of the standards. The spectra are shifted by 1.0 for clarity. The nebular contribution is manifested by the \ion{Ca}{ii} triplet, \ion{O}{i}\,8442 as well as numerous \ion{H}{i} Paschen series emission lines.}
    \label{fig:sp_type}
\end{figure}

If the M-giant is synchronized, a lower limit for the mass ratio is set by the ratio of its projected rotational velocity, $v\sin i$, to the orbital semi-amplitude.
We have derived the rotational velocity of the giant using the full width at half-maximum (FWHM) of a few unblended lines measured on our SALT spectra and compared with the synthetic spectrum of an M6.5 III star ( $T_{\rm eff}=3200 K$, $\log g = 0$). The adopted spectral type resulted from comparison of the low-resolution spectra of V3890 Sgr with those of M-type giant spectroscopic standards (Fig.~\ref{fig:sp_type}). 
The synthetic spectrum was from the BT-NextGen grid of theoretical spectra \citep{allard} that are available from `Theoretical spectra webserver' at the SVO Theoretical Model Services\footnote{http://svo2.cab.inta-csic.es/theory/newov2/index.php}.
As in the case of other symbiotic giants we adopted 2\,\kms and 3\,\kms for the micro and macro turbulence velocities.
The resulting mean  value, $v\sin i = 10.9\pm0.6$\,\kms, combined with $K_{\rm g}=16.62\pm0.38$ results in $q_{\rm min}=0.82\pm 0.09$ that is surprisingly close to the mass ratio derived from radial velocities. This indicates that the giant is filling or nearly filling its Roche Lobe (RL) - the actual $q$ should be then equal to $q_{\rm min}$, otherwise the measured $v\sin i$ is faster than the synchronized value. The first option is consistent with the apparent eccentricity of the M-giant orbit. Moreover, it is also easier to ensure the high mass transfer and accretion rate required by the activity and short nova outburst recurrence time of V3890 Sgr.

\subsection{Red giant and distance to V3890 S\lowercase{gr}}\label{distance}

\citet{munari2019} derived $E(B-V)=0.59$ and $d=4.5$ kpc from the equivalent widths of the interstellar (IS) \ion{K}{i} lines and DIBs. They also argue that this is a minimum value because there is no material in the direction of V3890 Sgr beyond $\sim 4.5$ kpc.
The {\it Gaia} parallax of V3890 Sgr is not very helpful because its DR2 value $\pi=0.1949\pm0.0943$\,mas, giving a distance $d=4.3^{+2.6}_{-1.3}$ kpc \citep{gaiadr2} which is very different from its EDR3 value $\pi=0.0484\pm0.0453$\,mas (Gaia Collaboration et al. 2020), with goodness of fit of $\sim 17$ in both cases indicating very poor fit.

The \ion{Na}{i} D1,2 interstellar line profiles consist of several components lines which indicate that the line-of-sight absorption consists of a number of discrete clouds each of which has some mean velocity. In particular, radial velocities with respect to the local standard of rest, $V_{\rm LSR}$, derived from \ion{Na}{i} D$_{1,2}$ lines indicate a range of mean cloud velocities of 2 to 80\kms (Fig.~\ref{profiles}).  If the velocity is due to Galactic rotation it can be used to derive a lower limit to the distance. In particular, the component with the maximum velocity, $V_{\rm LSR} = 80$\kms, indicates $d \ga 6.1$ kpc.

The velocity and intensity structure of \ion{Na}{i} D$_{1,2}$  in  V3890 Sgr is very similar  to  that  in the well studied O9.5 II-III star, HD 168941, located at d=5.8 kpc  \citep[see][fig. 4]{sembach1993}  of which the  line of sight (l,b=5.82,-6.31)  is about 3\degr away from V3890 Sgr.  Simultaneously, the systemic velocity of V3890 Sgr, $\gamma_{\rm LSR} = -82.3$\kms is inconsistent with Galactic rotation at any distance for its line-of-sight which means that V3890 Sgr does not belong to Galactic disk population but rather to the spheroidal Galactic bulge located beyond this interstellar cloud complex, and its distance should be $\ga 6$ kpc.

The maximum distance is set by the Roche lobe radius of the giant. The mass ratio, $q=0.78\pm0.05$, and  the binary separation, $a \sin i=2.010 \pm 0.065$\,AU, and $i \approx 69\degr$ (sec.~\ref{sp_orbits}) results in $R_{\rm RL}=166 \pm 8$\,\rsun\ which combined with  $T_{\rm eff}=3200 K$ gives the luminosity  $L_{\rm g} = 2600 \pm 250$\,\lsun\ and bolometric magnitude, $M_{\rm bol}=-3.79 \pm 0.10$ . The 2MASS magnitudes of V3890 Sgr, $J=9.832 \pm 0.022$, $H=8.772\pm0.025$, $K=8.257\pm0.023$, were measured near the photometric minimum ($\phi \sim 0$), and they should well represent the red giant magnitudes. The reddening-corrected $K_0=8.04\ \pm 0.02$ and $(J-K)_0=1.26 \pm 0.05$ combined with the corresponding bolometric correction, $BC(K)=2.88 \pm 0.05$ \citep{bessel1984} give the observed bolometric magnitude, $m_{\rm bol}=10.92 \pm 0.07$. The distance modulus is then $m-M = 14.71 \pm 0.17$ and $d=8.75^{+0.71}_{-0.66}$ kpc.

The pulsation period of the red giant, $P_{\rm pul}=104.5$ d, is consistent with $M_K=-5.9$  if due to the fundamental mode pulsation  \citep[C-sequence]{wood1999} which combined with $K_0=8.04$ indicate $d=6.1$ kpc. If it is due to the first overtone pulsation then $M_K=-6.8$ (B sequence) and $d=9.3$ kpc. 
The red giant mass, $M_{\rm g} \approx 1.05$\msun, and radius, $R_{\rm g} \approx 166$\rsun, imply the pulsation constant $Q = P_{\rm pul} (M_{\rm g}/M_{\sun})^{1/2}(R_{\rm g}/R_{\sun})^{-3/2} \approx 0.05$, consistent with the first overtone pulsation \citep[e.g.][]{fox1982}.

Finally, the maximum visual magnitude recorded during the 2019 nova outburst  of V3890 Sgr, $V_{\rm max}=8.30$ (JD 2\,458\,724.5; reddening-corrected $V_{\rm max,0}=6.47$)  results in the absolute magnitude $M_V \approx -8.3$ for the distance $d \approx 9$ kpc which is very similar to those of RNe in the Large Magellanic Cloud \citep[e.g.][]{shafter}.

We adopt the distance $d=9$ kpc for the rest of the paper.

\begin{figure}
\includegraphics[width=\columnwidth]{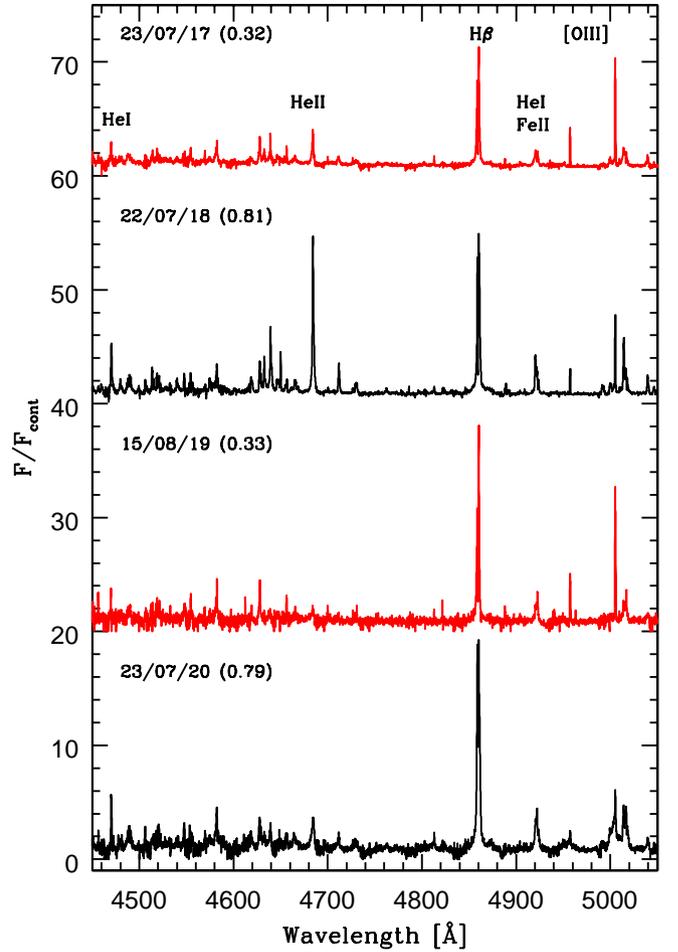}
\caption{Examples of SALT spectra of V3890 Sgr in blue range.}\label{spectra}
\end{figure}

\section{Inter-outburst activity}\label{activity}

In addition to the orbitally related changes and the M-giant pulsation the light curves of V3890 Sgr (Fig.~\ref{var}) show long-term variations on timescale of years. Similar behaviour has been reported for the other SyRNe. In particular, \citet{gromadzki2008} found light-variations with timescales of 1200-1800d in the inter-outburst visual light curves of RS Oph, \citet{mroz2014} reported quasi-periodic $\sim$2400d changes in the OGLE light curve of V745 Sco whereas \citet{ilkiewicz2016} discussed photometric and spectroscopic changes over $\sim 1000$ and $\sim$5000d timescales in T CrB. These changes resemble the classical Z And variability, which is believed to be due to unstable disc-accretion \citep{mik2003, sokoloski2006}. The inter-outburst activity of SyRNe could have the same origin as in the Z And-type symbiotic systems. 

\begin{figure*}
\resizebox{\hsize}{!}{\includegraphics{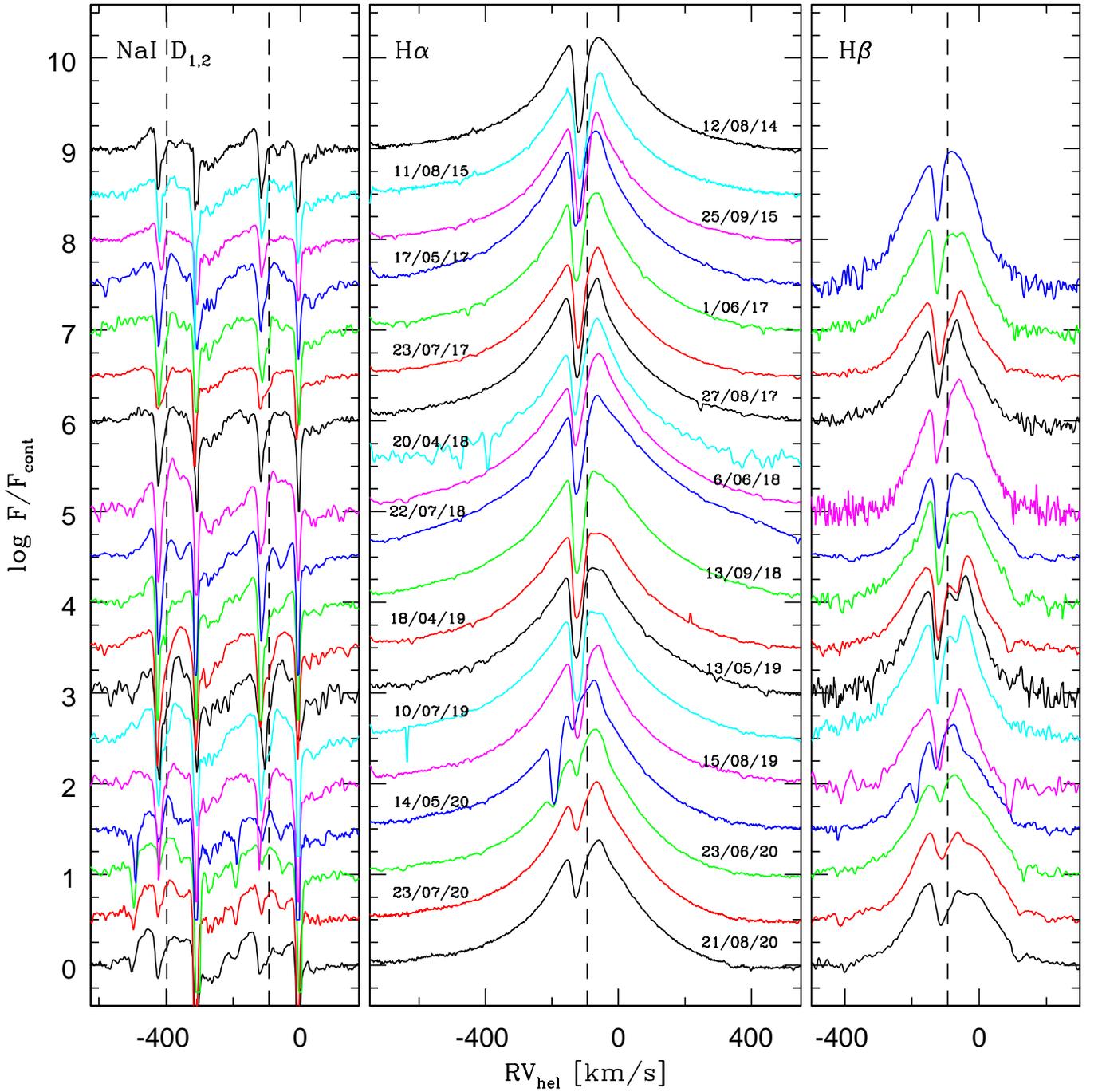}}
\caption{Emission line profiles in V3890 Sgr. Each profile was normalized to local continuum. The profiles are plotted in logarithmic flux scale and shifted vertically by 0.5 dex for better display. The dashed vertical lines mark the systemic velocity for each line.}\label{profiles}
\end{figure*}

Our high resolution SALT spectra taken prior to the nova outburst reveal remarkable activity associated with the photometric behaviour of V3890 Sgr. In particular, the spectra show a wealth of broad (FW$\sim$500-1000\kms) emission lines of \ion{H}{i}, \ion{He}{i}, \ion{Fe}{ii}, \ion{Na}{i}\,D and \ion{He}{ii} as well as narrow (FW$\sim$100\kms) forbidden lines of [\ion{O}{iii}], [\ion{O}{i}] and [\ion{N}{ii}] (Fig.~\ref{spectra}).
The broad emission line profiles (Fig.~\ref{profiles}) do not resemble CV accretion disc-like profiles. The H$\alpha$ wings are significantly broader than those in H$\beta$ indicating that scattering contributes to the line broadening. 
However, no trace of the Raman scattered \ion{O}{vi} is found even when \ion{He}{ii}\,4686 $\sim$ H$\beta$ (Fig.~\ref{spectra}).
 
The presence of strong \ion{He}{ii}\,4686 line is particularly remarkable because this line is rarely observed in quiescent SyRNe \citep[e.g.,][and references therein]{anupama1999, ilkiewicz2016}. In the case of V3890 Sgr, a strong \ion{He}{ii}\,4686 line was observed in 1981 \citep{williams1983} as well as in 1991 and 1992 \citep{williams1994}. Then a weak \ion{He}{ii}\,4686 line was again observed in April 1997 but absent in March 1998 \citep{anupama1999}. Similarly, the spectra taken in September 2011 (MJD 55821, $\phi$=0.43) and September 2012 (MJD 56186, $\phi$=0.91), respectively,  show the Balmer lines and lines of \ion{He}{i}, [\ion{O}{i}], and very weak \ion{He}{ii}\,4686\AA\,  \citep[see][fig. 11]{zemko2018}.

The \ion{He}{ii}\,4686 line was one of the strongest emission lines in the SALT spectra taken in 2016-2019 (spectra taken in 2014-2015 do not cover the blue region). Unfortunately, the SALT spectra are not flux-calibrated, thus we cannot estimate the emission line fluxes and the luminosity of the hot WD. However, we can still make some crude estimates comparing the V3890 Sgr emission line spectrum to those of symbiotic binaries with known parameters of their hot WD. To produce detectable \ion{He}{ii}\,4686 emission in a symbiotic system a hot, $\ga 10^5\,K$ and luminous, $\ga 100$\,\lsun\ source is required \citep[see, e.g.][]{mas97}. For a 1.35\,\msun WD this indicates a minimum accretion rate of $\ga 10^{-8}$\,\msun\, yr$^{-1}$. 

In fact, the intensity of \ion{He}{ii}\,4686 in relation to H$\beta$ has been at least $\ga 0.3$ (which requires the luminosity and accretion rate a few times higher than the minimum value) on all spectra but the one taken on August 15 2019 (JD 24558711.4), i.e. 12 days before the onset of the  TNR nova outburst (August 27.87 UT/JD 2458723.37). According to our orbital ephemeris the WD was then passing in front of the red giant ($\phi = 0.333$), and the dimming of \ion{He}{ii}\,4686 cannot be due to any eclipse effect. Moreover, the line was strong in 2017 (see Fig.~\ref{spectra}) when the system was in the same orbital phase which points to intrinsic variability of the temperature and brightness of the ionizing source.

The changes in the emission line intensity is clearly correlated with the photometric activity. The system was unusually faint in mid August 2019, $V \ga 16$, when the  \ion{He}{ii}\,4686 emission line was only marginally detected.
On the other hand, 
V3890 Sgr was particularly bright in 2018, with maximum $V_{\rm max} \approx 14.4$ and $I_{\rm c, max} \sim 12.2$ in July 2018 (Fig.~\ref{var}) when the spectrum showed very strong emission lines (\ion{He}{ii}\,4686 $\sim$ H$\beta$, EW(\ion{He}{ii}) $\approx$ 22\,\AA; Fig.~\ref{spectra}) simultaneously with enhanced blue continuum and cF-type absorption lines. 

The magnitude and colour of the additional source can be estimated by subtracting the red giant magnitudes ($V_{\rm g} \approx 16.5$ and $I_{\rm c,g} \approx 12.8$, respectively, corresponding to minimum observed magnitudes; Fig.~\ref{var}).
The resulting magnitudes corrected for the IS reddening, $E(B-V) \sim 0.6$ are  $V_0 \approx 12.7$, $I_{c,0} \approx 12.0$ and $(V-I_{c})_0 \approx 0.7$. The $V-I_{c}$ colour is consistent with a late F/early  G-type stellar photosphere\citep[see, e.g.,][]{mmu2014}. In the case of F/G-type giant, the bolometric correction $BC \approx 0$, the absolute magnitude $M_{\rm bol} \approx M_V \approx -2.1$,  and the luminosity $L \sim 530$\lsun. 
The colour is marginally consistent with  a {\it bf+ff} nebular emission, however, an unrealistically high volume emission measure of $\sim 5\times 10^{60} \rm cm^{-3}$ would be then required. 

It is tempting to associate the double-temperature structure of the active source in V3890 Sgr with the presence of an accretion disc. The blue continuum and the cF absorption would be then formed in the accretion disc and the material streaming towards the WD, respectively, whereas the high ionization emission lines would originate in low density material above and below the disc 
photoionized by high-energy boundary layer photons. 
The cF absorption lines apparently do not trace the WD orbit, and their maximum radial velocity is observed near $\phi \sim 0.8$ which points to their origin predominantly in the region where the stream encounters the disc (Fig.~\ref{rvel}).
Standard accretion disc theory predicts that half of the accretion energy is released in the disc and  another half in the boundary layer. If this is the case for V3890 Sgr, an accretion rate of $\sim 10^{-7}$\,\msun\, yr$^{-1}$ is required to power the luminosity of the blue continuum observed during the 2018 high state. Morever, the fact that the super-soft X-ray emission turns off after the nova outburst \citep{pag20,sin21} and it is not observed in quiescence indicates that the accretion rate remains below the nuclear burning range, $\dot{M}_{acc} \la 2.55 \times 10^{-7}$ \msun\,yr$^{-1}$ \citep{nomoto}, which is fully consistent with our estimates.

We now compare the WD mass and accretion rates derived from our observations 
with the values predicted by theoretical models. Detailed modeling of a dense grid of high mass accretion rates onto massive WDs found a tight correlation between the duration of a hydrogen flash and the mass of the underlying WD \citep{hil16}.
In the case of V3890 Sgr that duration, measured by the interval between when the eruption began and when supersoft 
x-ray emission stopped, is 26 days \citep{pag20}. Based on fig.5 and  eq.5 in \citet{hil16}, we estimate the WD mass of $\sim 1.38$\,\msun, in very good agreement with the dynamical mass derived in Section~\ref{sp_orbits}). 
For a given WD mass, \citet{hil16} showed that the time between nova eruptions is correlated with the mass accretion rate (see their fig. 4). For a 1.38\,\msun\ WD that accretion rate is $\sim 2 \times 10^{-8}$ \msun\,yr$^{-1}$, while for a 1.35\,\msun\ WD the rate is $\sim 4 \times 10^{-8}$ \msun\,yr$^{-1}$. These values are in good agreement with our estimates, $\dot{M}_{acc} \sim$ a\,few$\times 10^{-8}$--$10^{-7}$\,\msun\, yr$^{-1}$, based on the the quiescent system luminosity. 

The central part of the \ion{H}{i} and \ion{Na}{i}\,D profiles is dominated by complex absorption features. The main absorption component does not follow the orbital phase and it is blueshifted by $\sim 14$\,\kms\ relative to the systemic velocity. This absorption is presumably formed in circumstellar material. In addition, there is fainter absorption structure redshifted relative to the systemic velocity that seems to vary with the orbital phase, which we associate with the material flowing towards the WD. 

\begin{figure}
\includegraphics[width=\columnwidth]{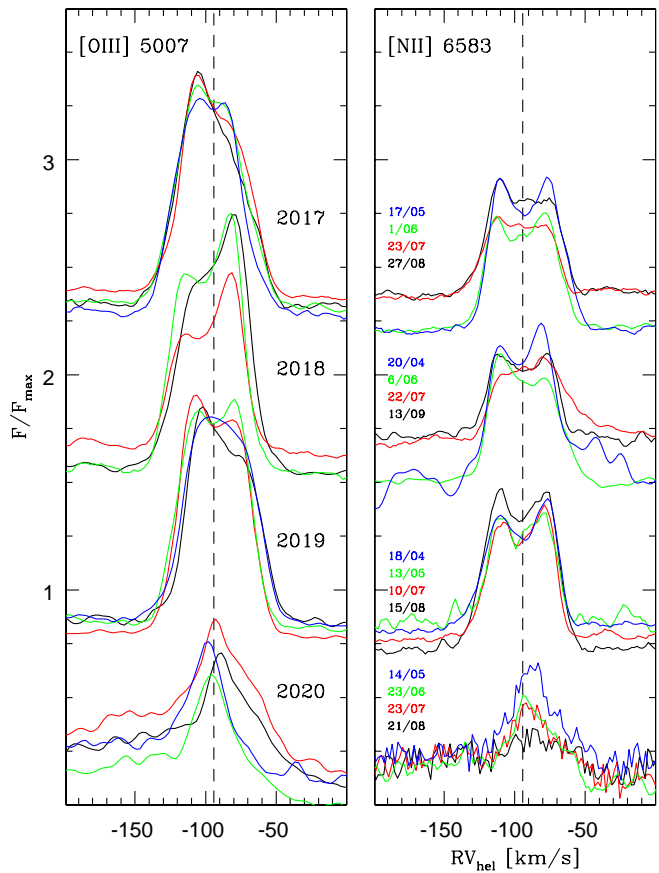}
\caption{Forbidden emission line profiles in V3890 Sgr. Each profile was normalized to maximum flux. The profiles from 2020 are stretched by a factor of 5  for clarity. The dashed vertical line marks the systemic velocity.}\label{forb}
\end{figure}

The strong and narrow forbidden lines of [\ion{O}{iii}], [\ion{O}{i}] and [\ion{N}{ii}] show double peaked profiles separated by $\sim 30$\,\kms, and centered at $\sim -94$\,\kms, i.e. the binary systemic velocity (Fig.~\ref{forb}). They remain relatively stable during 2014-2019. These narrow forbidden lines are likely formed in a shell surrounding the whole binary, presumably created by the previous nova outburst(s).

Simulations of the quiescent mass-transfer phase in the SyRN RS Oph \citep{booth} revealed complex and aspherical distribution of circumstellar material, in particular, a dense, outflow concentrated towards the orbital plane, and an accretion disc forms around the WD. These simulations also produced a polar accretion flow that can explain the broad wings of H$\alpha$ and other emission lines in quiescence. While the orbital parameters of V3890 Sgr are not identical with those of RS Oph, such dense equatorial flow may be common in symbiotic binaries. Very strong evidence that this is indeed the case is provided by recent direct imaging of the very innermost circumstellar region of R Aqr (a relatively wide symbiotic system with a Mira donor) which showed a strong focusing of the Mira wind in the equatorial plane, and material flowing towards the WD companion \citep{bujarrabal}. 

\citet{booth} also simulated the interaction between multiple novae and the circumstellar medium and found bipolar structures shaped by the interaction with the accretion disc and the wind: low-density and fast bipolar lobes and slow moving dense equatorial ring(s).
The \ion{H}{i} and \ion{Na}{i} D lines profiles in  2020, after the 2019 nova outburst, show very similar redshifted absorption features to those observed before, and in particular during  similar orbital phases in 2018. The cF absorption lines were also visible with similar velocities as in 2018. This indicates that the accretion flow has been fully restored after the nova outburst.
The only difference is the presence of an additional absorption feature blueshifted by  $\sim 180$ \kms\ with respect to the systemic velocity in \ion{H}{i} and \ion{Na}{i} D lines (Fig.~\ref{profiles}). This feature was very strong in May 2020, and in the case of \ion{H}{i} practically ceased by July 2020 whereas it has been  present in \ion{Na}{i} D until (at least) August. 
It seems reasonable to associate this component with the slow moving equatorial ring of swept-up wind material shown in the simulations of \citet{booth}.
The observed velocity $\sim -180$ \kms\ is consistent with the values of $\sim 100-200$ \kms\ resulting from these simulations.

The [\ion{O}{iii}], and [\ion{N}{ii}] profiles observed in 2020 have changed significantly in 2020, as there is only one narrow and faint component with EW a factor of $\sim 5$ lower than that in 2014-2019 (Fig.~\ref{forb}).

\begin{figure}
\includegraphics[width=\columnwidth]{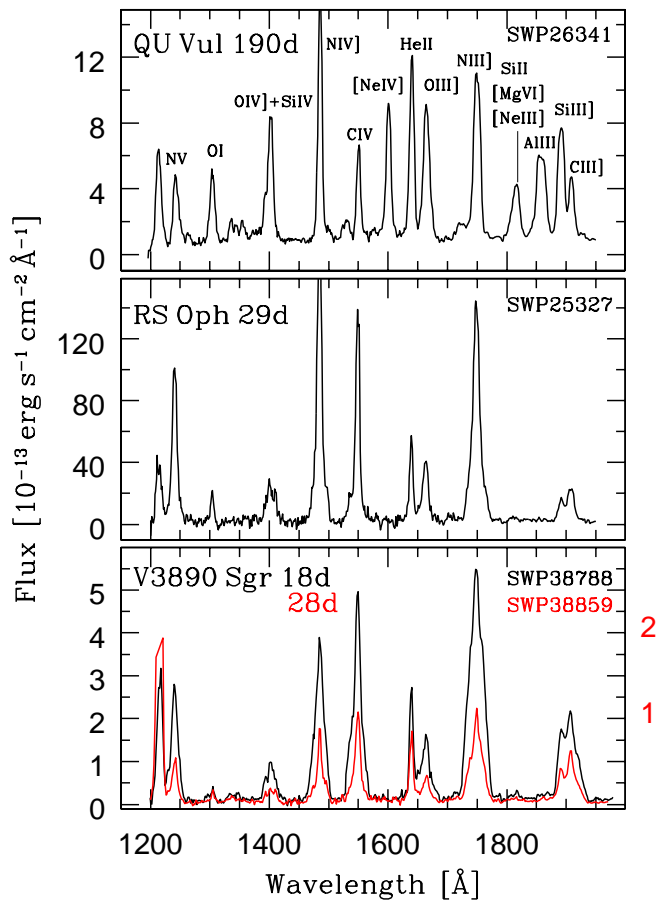}
\caption{ Comparison of {\it IUE} low-resolution spectra of V3890 Sgr, the symbiotic CO recurrent nova RS Oph and the ONe nova QU Vul. The spectra were taken during similar phases of the nova evolution.}\label{iue}
\end{figure}

\section {V3890 Sgr as a possible progenitor of SN I\lowercase{a}}\label{snia}

We now turn to the question of linking SyRNe, and in particular V3890 Sgr, to single-degenerate progenitors of SN Ia.
Because our results indicate the WD is very massive, $\goa 1.35$\,\msun\ it is natural to ask whether it hosts a CO WD (as only CO WD can give rise to SN Ia) that has grown to its present value due to accretion.

By comparing UV spectra of the SyRN RS Oph with those of novae erupting on CO  and ONe WDs, \citet{MS2017} demonstrated that the RS Oph WD is clearly made of CO. Fig.~\ref{iue} compares two low-resolution {\it IUE} spectra of V3890 Sgr taken on the 18th and 28th days after the onset of the 1990 outburst with the spectra of RS Oph and QU Vul - the ONe nova from \citet{MS2017}. The spectra of both SyRNe are practically identical, and remarkably different from that of the ONe nova which indicates novae erupting on CO WDs in both cases.
This appears to be in contention with Chandra spectroscopy of V3890 Sgr on the 7th day after the 2019 nova outburst onset that indicates a CO WD based on the ratios of Al, Mg and Ne line fluxes relative to the Si and Fe fluxes \citep{orio2020}. 
V3890 Sgr is thus  the second SyRN with a confirmed massive CO WD that could eventually give rise to a SNIa explosion.

The estimated radius of the giant, $R_{\rm g} \approx 166$\,\rsun\, (sec.~\ref{distance}) indicates a core mass $\sim 0.5$\,\msun\, \citep{rapp1995}. This means that $\sim 0.5$\,\msun\,  is still stored in the giant's envelope, much more than  $\sim 0.05-0.1$ \msun\, needed to grow the WD to the Chadrasekhar limit. 
Using the expression by \citet{kudritzki} we estimate the wind mass loss of $2.4 \times10^{-7}$\,\msun\, yr$^{-1}$.  Similarly, a mass transfer rate of $5\times10^{-7}$\,\msun\, yr$^{-1}$ results from \citet{ritter}'s formula. The expected lifetime of the red giant would be then $\sim 10^6$ yrs, long enough to deposit and retain the missing $\sim 0.05-0.1$ \msun\, on the WD even if only a fraction of the red giant wind will be accreted.

\section{Conclusions}\label{conclusions}

Based on nineteen SALT spectra of V3890 Sgr obtained in 2014-2020 combined with 20 years of $V,I_{C}$ photometry we have determined the orbital period to be 747.6 days which definitely rules out the 519.7-day period reported by \citet{schaefer2009}. Our double-line spectroscopic orbits indicate 
the mass ratio $q=M_{\rm g}/M_{\rm WD} = 0.78 \pm 0.05$, 
very close to the minimum mass ratio, $q_{\rm min}=0.82\pm 0.09$, set by the ratio of rotational velocity to the orbital semi-amplitude of the red giant as expected in the case of the giant filling (or nearly filling) its Roche lobe.
The presence of a RL-filling giant is also consistent with the eccentric orbit solution tied to the line of sight.
The most likely orbital solution for the component masses are $M_{\rm WD} \approx 1.35\pm0.13$\,\msun\, and  $M_{\rm g} \approx 1.05 \pm 0.11$\,\msun, while the orbit inclination is $\approx 67-69\degr$.

The distance to V3890 Sgr set by the RL radius of the giant, $d \approx 9$ kpc, is consistent with the value indicated by the giant pulsation characteristics, and the minimum value set by the interstellar \ion{Na}{i}\,D$_{1,2}$ lines. Moreover, at this distance, the absolute outburst magnitude of V3890 Sgr is very similar to those of RNe in the Large Magellanic Cloud.

V3890 Sgr shows remarkable photometric and spectroscopic activity between the nova eruptions with timescales similar to those observed in the SyRNe T CrB and RS Oph and Z And-type symbiotic systems. 
The  active source has a double-temperature structure which we have associated with the presence of an accretion disc. The activity would be then caused by changes in the accretion rate. 
The radial velocity pattern in the blue cF absorption spectrum is consistent with formation in the material streaming from the giant presumably in the region where the stream encounters the disc.

The double-peaked narrow forbidden line profiles remain relatively stable during 2014-2019, and  they are likely formed in a shell surrounding the whole binary, presumably created by the previous nova outburst(s).
The spectra taken in 2020 indicate that the accretion flow has been fully restored after the nova outburst, however, the outer circumstellar regions do not.

Finally, there is strong evidence that V3890 Sgr contains a CO WD accreting at a high, $\sim$ a\,few$\times 10^{-8}$--$10^{-7}$\,\msun\,yr$^{-1}$,  rate. The WD is growing in mass, and should give rise to a SNIa event within $\la 10^6$ yrs - the expected lifetime of the red giant.

\section*{Acknowledgements}
This research has been partly financed by the Polish National Science Centre (NCN) grants OPUS 2017/27/B/ST9/01940 and MAESTRO 2015/18/A/ST9/00746.
MOH is financed by the Polish NCN grant SONATINA 2019/32/C/ST9/00577.
The paper is based on spectroscopic observations made with the Southern 
African Large Telescope (SALT) under programmes 2014-1-POL\_RSA\_AMNH-001, 2015-1-SCI-028, 2017-1-SCI-038, 2017-2-SCI-017 and
2018-1-MLT-005 (PI: J. Miko{\l}ajewska). Polish participation in SALT is 
funded by grant No. MNiSW DIR/WK/2016/07.
The OGLE project has received funding from the Polish NCN grant MAESTRO 2014/14/A/ST9/00121 to AU.

\section*{Data availability statement}
SALT spectra taken prior to 2018 are publicly available. The other data underlying this paper are available on reasonable request to the corresponding author.







\bsp	
\label{lastpage}
\end{document}